\documentclass[reprint, amsmath,amssymb,aps,pra,longbibliography,superscriptaddress]{revtex4-2}
\usepackage{graphicx}
\usepackage{dcolumn}
\usepackage{bm}
\usepackage{float}

\usepackage[dvipsnames]{xcolor}
\usepackage[colorlinks=true,
linkcolor=blue, 
citecolor=purple,   
urlcolor=Mahogany       
]{hyperref}

\begin{document}

\author{Hao Wang}
\email{wang\_hao1@tju.edu.cn}
\address{State Key Laboratory of Engines, Tianjin University, Tianjin 300072, China}	
\author{Li Zhao}
\address{State Key Laboratory of Engines, Tianjin University, Tianjin 300072, China}
\author{Shuai Deng}
\address{State Key Laboratory of Engines, Tianjin University, Tianjin 300072, China}
\author{Yu-Han Ma}
\email{yhma@bnu.edu.cn}
\address{School of Physics and Astronomy, Beijing Normal University, Beijing, 100875, China}
\address{Key Laboratory of Multiscale Spin Physics (Beijing Normal University), Ministry of Education, Beijing 100875, China}

	\date{\today}
	
	\title{Thermodynamic Geometry of Relaxation}

	\begin{abstract}
    While the geometry of equilibrium states and driven non-equilibrium processes is clearly understood, a geometric description for relaxation towards equilibrium is still lacking. Here, we propose a thermo-geometric measure based on the Rayleigh quotient, reformulating relaxation as a fundamental competition between entropic stiffness and frictional dissipation. Taking a van der Waals gas with two dissipation channels as an example, we explicitly demonstrate its relaxation landscape. Particularly, we find that upon approaching the critical temperature $T_c$, the slow-mode relaxation rate vanishes linearly as $\lambda_s \propto (T-T_c)/T_c$, indicating critical slowing down. This study completes the thermodynamic geometry framework, providing a general tool for characterizing the relaxation dynamics of complex systems.
\end{abstract}

\maketitle

\addtocontents{toc}{\protect\setcounter{tocdepth}{-2}}

\textit{Introduction}.---The thermodynamic geometry framework~\cite{weinholdMetricGeometryEquilibrium1975,ruppeinerThermodynamicsRiemannianGeometric1979b,schloglThermodynamicMetricStochastic1985,quevedoGeometrothermodynamics2007a} has advanced our understanding of complex systems by formulating the description of equilibrium properties in the rigorous language of Riemannian manifolds. By mapping phase transition singularities to curvature divergences~\cite{georgeRiemannianGeometryThermodynamic1995,Janyszek1990,ThermodynamicGeometryofSupercooledWater2015} and connecting macroscopic distances to microscopic fluctuations~\cite{ruppeinerThermodynamicsRiemannianGeometric1979b}, this geometric perspective provides fundamental structural insights. Recently, stochastic thermodynamics and optimal transport theory have extended this picture to nonequilibrium processes, revealing that the most efficient protocol (associated with minimum dissipation) connecting two states follows a geodesic path, whether mapped onto the parameter space of friction tensors or the statistical manifold of the Wasserstein metric~\cite{PhysRevLett.51.1127,diosiThermodynamicLengthTime1996,crooksMeasuringThermodynamicLength2007,Aurell2011,SivakCrook2012,vanvuGeometricalBoundsIrreversibility2021,kimInformationGeometryFluctuations2021,LiGengGeodesicPath2022,chenGeodesicLowerBound2023,zhongLinearResponseEquivalence2024}.

Despite these advances, a fundamental gap persists. While current geometric frameworks primarily focus on driven processes controlled by external protocols~\cite{Aurell2011,SivakCrook2012,abiusoOptimalCyclesLowDissipation2020a,vanvuGeometricalBoundsIrreversibility2021,kimInformationGeometryFluctuations2021,frimOptimalFinitetimeBrownian2022,2022Ma,LiGengGeodesicPath2022,chenGeodesicLowerBound2023,zhongLinearResponseEquivalence2024,2024Zhao}, a universal theory governing the intrinsic dynamics of \textit{relaxation} (the un-driven evolution by which a system naturally seeks equilibrium within a dissipative environment) remains absent. Such spontaneous equilibration is the most common thermodynamic process in nature and engineering; however, it frequently conceals profound dynamical complexities. For systems featuring coupled dissipation channels, this complexity inherently manifests as disparate timescales, raising a pivotal question of whether this temporal hierarchy, and specifically the anomalous decay of relaxation rates near phase transitions, is directly encoded in the underlying topography of the state space. Physically, this process is not uniform motion along a geodesic but rather an inherent competition between entropic forces and kinetic resistance~\cite{Onsager1931a,Callen:450289}. Although established analytical frameworks~\cite{eigenMethodsInvestigationIonic1954,gorbanConstructiveMethodsInvariant2004,gorbanInvariantManifoldsPhysical2005,transtrumGeometryNonlinearLeast2011,transtrumPerspectiveSloppinessEmergent2015} successfully capture these multiscale phenomena via dynamic approaches, the geometric origin of these timescales and their critical behavior has not yet been unveiled.

In this Letter, we bridge this gap by proposing a unified geometric measure for relaxation processes. We formulate the intrinsic relaxation rates $\lambda$ and their corresponding principal axes $\bm{v}$ directly through the thermodynamic Rayleigh quotient
\begin{equation}\label{eq:Rayleigh_quotient}
\lambda=\mathcal{R}(\bm{v})=\frac{\bm{v}^\top \bm{g}\bm{v}}{\bm{v}^\top \bm{a} \bm{v}}=\frac{||\bm{v}||_{\bm{g}}^2}{||\bm{v}||_{\bm{a}}^2},
\end{equation}
which intrinsically integrates two fundamental thermodynamic tensors: the Ruppeiner metric $\bm{g}$ (representing static entropic stiffness) and the Onsager matrix $\bm{a}$ (representing kinetic dissipation). Serving as a coordinate-independent geometric probe, this quotient identifies collective modes by quantifying the local competition between restorative driving and viscous entropy production. Applying this framework to the van der Waals fluid, we unveil the geometric origin of critical slowing down. We demonstrate that near criticality, temporal scaling laws emerge from an asymptotic mismatch between the two tensors. Specifically, the entropic metric undergoes a geometric collapse that traps the slow mode within its null space.

\begin{figure*}
	\centering
	\includegraphics[width=1\linewidth]{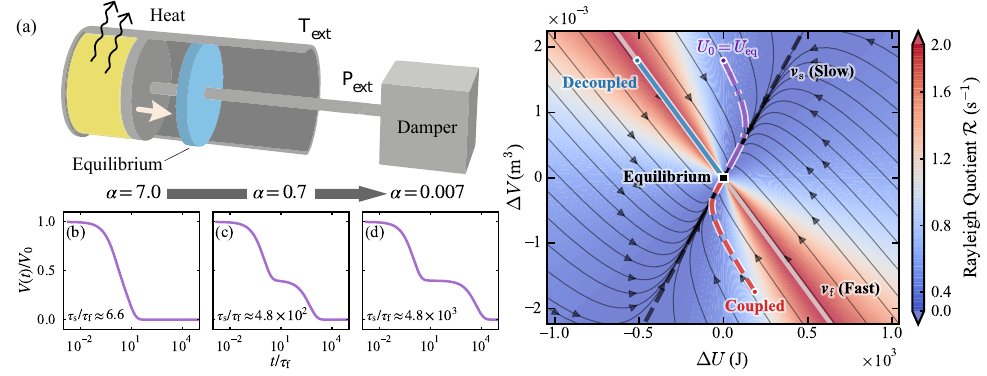}
	\caption{(a) Schematic of a damped piston-cylinder system enclosing a van der Waals fluid. (b)-(d) Temporal evolution under decreasing thermal transport parameter $\alpha$, exhibiting a distinct two-stage decay with a extended plateau. (e) Landscape of the Rayleigh quotient $\mathcal{R}(\bm{v})$ in the thermodynamic state space. Solid gray ($\bm{v}_\mathrm{f}$) and dashed black ($\bm{v}_\mathrm{s}$) lines denote the fast and slow principal axes. Representative trajectories depict purely decoupled (solid blue), generically coupled (dashed red), and geometrically constrained (dash-dotted purple, corresponding to the two-stage decay in panel b) relaxation dynamics. Fixed parameters: $T_{\text{ext}} = 300\,$K, $P_{\text{ext}} = 2 \times 10^5\,$Pa, $\beta = 6 \times 10^{-8}\,\text{m}^3/(\text{Pa}\cdot\text{s})$, and $\alpha = 7\,$W/K (for panel e).}
	\label{fig:Fig_1}
\end{figure*}

\textit{Geometric Variational Framework.}---The physical necessity of this bi-metric formulation stems from the underlying mechanics of relaxation. In multivariable coupled systems, the evolution path is governed by the instantaneous balance between thermodynamic driving forces and dissipation. Because a single geometric tensor is insufficient to capture this interplay, we integrate the equilibrium entropic landscape ($\bm{g}$)~\cite{ruppeinerThermodynamicsRiemannianGeometric1979b} and the generalized macroscopic friction ($\bm{a}$)~\cite{Sprikl1995,SivakCrook2012} into a unified relaxation equation ~\cite{SM}
\begin{equation}\label{eq:Phenomenological_equation}
\dot{\bm{x}}=\bm{a}^{-1}\bm{g}\bm{x}\equiv-\bm{\Gamma} \bm{x}.
\end{equation}
Here, the rate tensor $\bm{\Gamma}$ acts geometrically to distort the static stiffness of $\bm{g}$ according to the dissipative background of $\bm{a}$. Constrained by macroscopic thermodynamic consistency (Maxwell relations)~\cite{Callen:450289} and microscopic reversibility (Onsager reciprocity)~\cite{Onsager1931a}, $\bm{\Gamma}$ satisfies a generalized self-adjointness with respect to $\bm{g}$, namely, $\bm{\Gamma}^\top \bm{g} = \bm{g}\bm{\Gamma}$. This crucial symmetry guarantees real eigenvalues and ensures the simultaneous diagonalization of both metrics~\cite{Horn_Johnson_2012,gorbanInvariantManifoldsPhysical2005}. Consequently, the eigenbasis of $\bm{\Gamma}$ spans a set of intrinsic principal axes wherein multivariable relaxation completely decouples into independent exponential decay modes. (Incorporating the energy-representation Weinhold metric naturally generalizes this relation to a covariant four-metric framework~\cite{SM}.)

While algebraic diagonalization successfully isolates these dynamical modes, it obscures their profound thermodynamic origins. To explicitly expose the inherent competition between restorative driving and viscous drag, we recast the macroscopic dynamics into a geometric variational problem. Dictated by Onsager's linear response, the instantaneous evolution reflects a strict force balance between the restorative gradient $\bm{g}\bm{x}$ and the viscous drag $\bm{a}\dot{\bm{x}}$~\cite{gyarmatiNonequilibriumThermodynamics1970}. Integrating this balance over the entire relaxation trajectory yields a thermodynamic variational functional~\cite{finlaysonMethodWeightedResiduals1972}:
\begin{equation}\label{eq:Galerkin_weighted_residual}
\Pi = \int^\infty_0 \bm{x}^\top (\bm{a}\dot{\bm{x}}+\bm{g}\bm{x})dt.
\end{equation}
By substituting an arbitrary exponential trial path $\bm{x}=\bm{u}e^{-\lambda t}$, the Ritz stationarity condition $\partial \Pi/\partial \bm{u}=\bm{0}$ aligns the optimal evolution direction exactly with an intrinsic principal axis $\bm{v}$. This variational formulation identifies the intrinsic relaxation rate $\lambda$ as the stationary value of the thermodynamic Rayleigh quotient $\mathcal{R}(\bm{v})$ defined in Eq.~\eqref{eq:Rayleigh_quotient}~\cite{SM}. Geometrically, $\lambda$ emerges as a coordinate-independent invariant, defined in the local tangent space by the ratio of two squared Riemannian lengths: $||\bm{v}||_{\bm{g}}^2 / ||\bm{v}||_{\bm{a}}^2$. This ratio encapsulates the fundamental physical competition between the driving potential (the static geometry, $\bm{v}^\top\bm{g}\bm{v}$) and the entropy production cost (the dissipative geometry, $\bm{v}^\top\bm{a}\bm{v}$). Consequently, $\mathcal{R}(\bm{v})$ acts as a powerful dynamic probe in the state space: fast modes locate at its maxima, representing the most kinetically efficient paths for releasing potential energy, whereas slow modes emerge at its minima, physically dictating the dynamic bottlenecks of macroscopic relaxation.

\textit{The Landscape of Relaxation}.---To demonstrate our general geometric framework with a concrete example, we consider a damped piston-cylinder system containing a real gas, as shown in Fig.~\ref{fig:Fig_1}(a). The equilibrium state of the gas is governed by the van der Waals equation of state, $P = RT/(V-b) - a/V^2$, where $V$, $P$, and $T$ denote the volume, pressure, and temperature of the gas, respectively; $R$ is the gas constant, and $a$ and $b$ are species-specific parameters. In contrast to standard finite-time thermodynamic models that typically treat heat transfer as the sole source of dissipation~\cite{chenExtrapolatingThermodynamicLength2021,PhysRevLett.125.210601}, we characterize the system's relaxation via two distinct channels: thermal conduction and mechanical damping. Governed by Newton's law of cooling and an overdamped piston subject to linear viscous drag, the macroscopic fluxes are naturally driven by their conjugate thermodynamic forces~\cite{SM}: the heat flux $dQ/dt = \alpha(T_{\text{ext}} - T)$ and the volumetric rate $dV/dt = \beta(P-P_{\text{ext}})$, where $\alpha$ and $\beta$ denote the respective transport coefficients. By substituting these macroscopic fluxes into the first law of thermodynamics ($dU = dQ - P dV$), we obtain a set of coupled relaxation equations that strictly govern the state-space dynamics. Geometrically, this intrinsic evolution is characterized by two fundamental tensors. The first is the static Ruppeiner metric~\cite{georgeRiemannianGeometryThermodynamic1995,SM}, which is determined directly by the equation of state as
\begin{equation}\label{eq:matric1}
\bm{g}= \frac{1}{T^2} 
\begin{pmatrix} 
C_V^{-1} & \chi \\ 
\chi & C_V\chi^2 - T \Xi 
\end{pmatrix},
\end{equation}
where $\Xi \equiv (\partial P/\partial V)_T= -RT(V-b)^{-2} + 2aV^{-3}$ and $\chi=-a/(C_V V^2)$. On the dynamic side, the dissipation metric~\cite{SM}
\begin{equation}\label{eq:matric2}
\bm{a}= \frac{1}{\alpha T^2} 
\begin{pmatrix} 
1 & P \\ 
P & \alpha T\beta^{-1} + P^2 
\end{pmatrix}, 
\end{equation}
is constructed from the coupled relaxation equations, where the boundary-work interplay naturally induces its non-zero off-diagonal elements. Consequently, the intrinsic relaxation rates $\lambda_{\rm{f(s)}}$ emerge naturally as the eigenvalues of the rate matrix $\bm{\Gamma}=\bm{a}^{-1}\bm{g}$, which in turn directly define the fast and slow characteristic timescales $\tau_{\rm{f(s)}} \equiv \lambda_{\rm{f(s)}}^{-1}$. 

By tuning the thermal transport parameter $\alpha$, we directly manipulate the system's temporal hierarchy. To unveil this kinetic asymmetry, Figs.~\ref{fig:Fig_1}(b-d) plot the macroscopic volume evolution against the dimensionless time $t/\tau_\mathrm{f}$. Decreasing $\alpha$ severely restricts thermal transport, drastically broadening the timescale separation. As the intrinsic timescale ratio $\tau_\mathrm{s}/\tau_\mathrm{f}$ escalates from approximately $6.6$ to $4.8 \times 10^3$, the dynamics phenomenologically split into a distinct two-stage decay. Under this temporal rescaling, the initial rapid mechanical decays perfectly collapse onto a single curve, exposing a progressively extended quasi-stationary plateau dictated entirely by the slow thermal bottleneck. 

Applying the metrics in Eqs.~\eqref{eq:matric1} and \eqref{eq:matric2} to Eq.~\eqref{eq:Rayleigh_quotient} maps out the Rayleigh quotient landscape $\mathcal{R}(\bm{v})$ in the $(\Delta U, \Delta V)$ state space, as shown in Fig.~\ref{fig:Fig_1}(e). The extremal ridges and valleys of this landscape dictate the intrinsic coordinate system: the fast ($\bm{v}_\mathrm{f}$, solid gray) and slow ($\bm{v}_\mathrm{s}$, dashed black) principal axes. While they appear non-orthogonal in the standard Cartesian frame, they are strictly orthogonal with respect to the thermodynamic metrics. The geometrically constrained trajectory, which corresponds to the distinct two-stage decay in Fig.~\ref{fig:Fig_1}(b) and is denoted by the dash-dotted purple line, translates the time-domain evolution onto this topographic terrain. Initially, the system descends along the high-$\mathcal{R}$ ridge, rapidly dissipating its fast-mode components (directly mirroring the initial volume drop). It then sharply turns into the valley of the slow manifold. This spatial confinement within the geometric valley physically governs the extended quasi-stationary phase. A generic coupled trajectory (dashed red) follows a similar hierarchical route, whereas paths initiated precisely along an eigendirection (solid blue) evolve exclusively within a single mode, completely bypassing the two-stage crossover. This geometric confinement to the slow manifold naturally generalizes to broader multiscale irreversible processes. Driven by the strong anisotropy between the static restoring forces ($\bm{g}$) and the kinetic resistance ($\bm{a}$), high-dimensional systems inherently seek paths of steepest restorative gradients per unit dissipation cost. This rapid initial descent effectively collapses their long-time evolution onto a lower-dimensional submanifold. While established invariant manifold theories algebraically capture such dimensionality reduction via spectral gaps~\cite{gorbanInvariantManifoldsPhysical2005}, our bi-metric framework unveils its pure thermodynamic origin: macroscopic timescale separation emerges directly from the fundamental geometric competition between these two intrinsic metrics.

\begin{figure}
	\centering
	\includegraphics[width=1\linewidth]{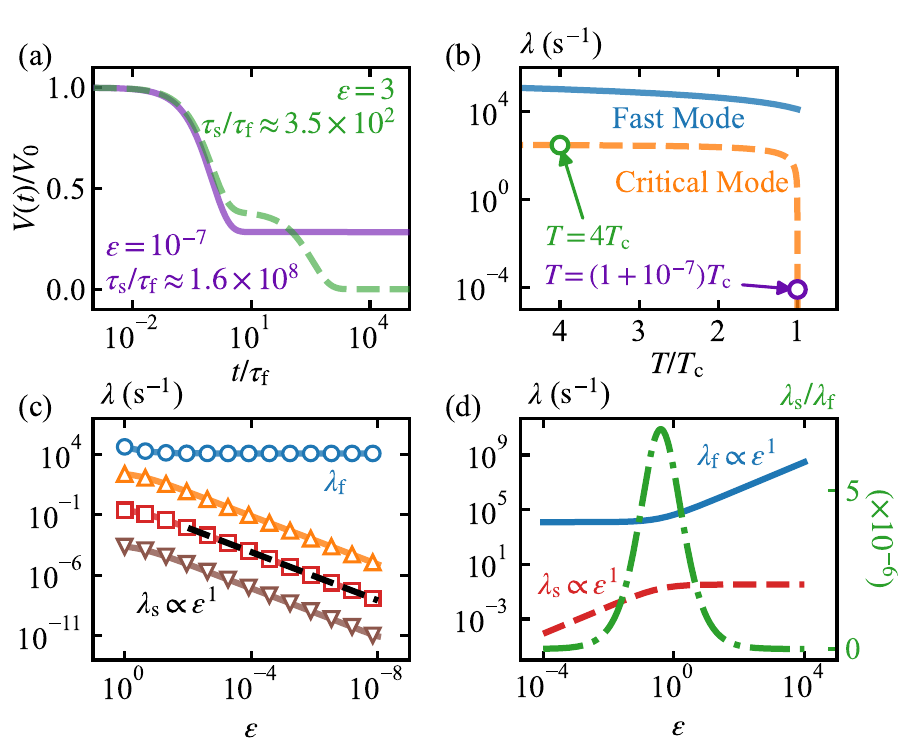}
    \caption{(a) Relaxation of the gas volume for $\epsilon = 10^{-7}$ (solid purple) and $3$ (dashed green) on the critical isochore. (b) Evolution of the fast ($\lambda_{\mathrm{f}}$) and slow ($\lambda_{\mathrm{s}}$) relaxation rates approaching $T_{\mathrm{c}}$ along the critical isochore. Open circles denote the states in panel (a). (c) The linear scaling law (black dashed line) holds for $10^{3}\alpha$ (orange triangles, $\alpha$ (red squares), and $10^{-3}\alpha$ (brown inverted triangles), with blue circles for the fast mode $\lambda_{\mathrm{f}}$. (d) Global dynamic landscape from the near-critical region to the macroscopic high-temperature extreme ($\epsilon \to \infty$). The timescale ratio $\lambda_{\mathrm{s}}/\lambda_{\mathrm{f}}$ (green dash-dotted curve) exhibits a peak. Here, $\alpha = 7\,$W/K, and other parameters remain as in Fig.~\ref{fig:Fig_1}.}
	\label{fig:scaling}
\end{figure}

\textit{Critical Slowing Down and Asymptotic Dynamics.---}To highlight the versatility of the bi-metric framework, we investigate the asymptotic geometric behaviors underlying critical phenomena and macroscopic limits. As the system approaches the critical point $T_{\mathrm{c}}$ ($\epsilon \to 0$, where $\epsilon \equiv T/T_{\mathrm{c}} - 1$), the separation of timescales progressively diverges. As illustrated by the contrasting volume relaxations in Fig.~\ref{fig:scaling}(a), $\tau_\mathrm{s}/\tau_\mathrm{f}$ surges to $\sim 10^8$ for the near-critical state, trapping the evolution into an extended quasi-stationary plateau (dynamic stagnation). This retardation is driven by a vanishing slow-mode rate ($\lambda_{\mathrm{s}} \to 0$) alongside a finite fast mode ($\lambda_{\mathrm{f}}$) shown in Fig.~\ref{fig:scaling}(b). Strikingly, varying the kinetic parameter $\alpha$ across six orders of magnitude only rescales the absolute rates but leaves the linear scaling law $\lambda_{\mathrm{s}} \propto \epsilon^1$ strictly preserved, as shown in Fig.~\ref{fig:scaling}(c). This robustness demonstrates that critical slowing down is not merely a consequence of specific transport coefficients, but rather an intrinsic geometric property of the thermodynamic state space. Confirming its non-trivial nature, this geometric scaling is analytically absent in non-interacting ideal gases~\cite{SM}. Thus, our framework allows critical scaling laws to emerge naturally from the metric structure itself, without requiring predefined order parameters.

Geometrically, critical slowing down originates from the vanishing isothermal stiffness, $\Xi = (\partial P/\partial V)_T \to 0$, a signature of mean-field criticality~\cite{landau1980fluctuation}. This vanishing stiffness triggers a geometric collapse, reducing the rank of the static metric $\bm{g}$~\cite{SM}. At $T_{\mathrm{c}}$, a profound asymptotic mismatch occurs: while the dissipative metric $\bm{a}$ remains regular, $\bm{g}$ degenerates into a rank-1 outer product
\begin{equation}\label{eq:g_rank_reduction}
\bm{g}(T_{\mathrm{c}}) \propto \begin{pmatrix} 1 \\ -\pi_{\mathrm{c}} \end{pmatrix} (1, -\pi_{\mathrm{c}}),
\end{equation}
where $\pi_{\mathrm{c}} = (\partial U/\partial V)_{T_{\mathrm{c}}}$ denotes the internal pressure at criticality. Dictated by the variational principle in Eq.~\eqref{eq:Rayleigh_quotient}, the slow mode $\bm{v}_{\mathrm{s}} = (\Delta U, \Delta V)^\top$ minimizes the restorative penalty by aligning with the null space of this degenerate metric. Combined with the thermodynamic identity $dU = C_V dT + (\partial U/\partial V)_T dV$, this geometric constraint, $\Delta U - \pi_{\mathrm{c}} \Delta V = 0$, uniquely yields $C_V \Delta T = 0$. Consequently, the slow mode is geometrically confined to the isothermal manifold. Because the dynamics of the system is locked onto this specific subspace, the slow relaxation rate is governed entirely by the vanishing stiffness $\Xi$, directly yielding the scaling law $\lambda_{\mathrm{s}} \propto |\Xi| \propto \epsilon^1$.

Beyond the critical point, we geometrically classify the temporal hierarchy across other macroscopic limits of the van der Waals fluid. By evaluating the determinant of the relaxation rate matrix $\bm{\Gamma}= \bm{a}^{-1}\bm{g}$, the off-diagonal thermomechanical cross-couplings perfectly cancel out, yielding a fundamental global rate identity~\cite{SM}:
\begin{equation}\label{eq:global_invariant}
\lambda_{\mathrm{f}} \lambda_{\mathrm{s}} = \det(\bm{a}^{-1})\det(\bm{g}) = -\frac{\alpha \beta}{C_V} \Xi.
\end{equation}
This scalar identity elegantly links the global timescale hierarchy to the macroscopic stiffness. For instance, in the dilute limit ($P \to 0$), $\bm{g}$ undergoes another rank reduction. Here, manifold locking asymptotically confines the fast mode to the isochoric manifold ($\Delta V = 0$) and the slow mode to the isoenergetic/isothermal manifold ($\Delta U = 0$), resulting in the scaling $\lambda_{\mathrm{s}} \propto P^2$~\cite{SM}. Both the critical ($\lambda_{\mathrm{s}} \propto \epsilon^1$) and dilute ($\lambda_{\mathrm{s}} \propto P^2$) scaling laws share the same physical origin: dynamical trapping within the flat directions of a flattened entropic potential landscape. Conversely, at macroscopic extremes ($\epsilon \to \infty$ or $P \to \infty$), the geometric mechanism operates in reverse. Instead of a flattened potential, the mechanical resistance ($\propto P^2$) diverges, dominating the dissipation metric $\bm{a}$ and reducing it to a rank-1 singular tensor $\bm{a}^{\mathrm{div}} \propto (1, P)^\top (1, P)$~\cite{SM}. Because any motion outside this null space experiences infinite thermodynamic friction( with infinite dissipation), the fast mode $\bm{v}_{\mathrm{f}}$ is geometrically confined to the null space of $\bm{a}^{\mathrm{div}}$. This yields $\Delta U + P\Delta V = \Delta Q = 0$, indicating that the fast mode is locked onto the adiabatic manifold with diverging rates ($\lambda_{\mathrm{f}} \propto \epsilon^1$ and $\propto P^2$). Concurrently, governed by the asymptotic eigenstructure and the global invariant given by Eq.~\eqref{eq:global_invariant}, the slow mode is locked onto the isobaric manifold ($\Delta P \approx 0$), and its rate saturates at a finite plateau $\lambda_{\mathrm{s}} \to \alpha/C_P$, clearly visible in the high-temperature limit of Fig.~\ref{fig:scaling}(d). Therefore, while slow-mode asymptotics arise from a flattened stiffness, fast-mode divergences are dictated by dynamical confinement within a singular dissipation landscape.

Furthermore, our framework provides insights into the global crossover of timescales. The peak of the timescale ratio $\lambda_{\mathrm{s}}/\lambda_{\mathrm{f}}$ in Fig.~\ref{fig:scaling}(d) signifies a state of minimum timescale separation. Here, the intrinsic anisotropies of $\bm{g}$ and $\bm{a}$ optimally counterbalance, creating an isotropic landscape that blurs the distinction between fast and slow manifolds and maximizes thermomechanical dissipation coupling. Finally, the bi-metric framework autonomously exposes structural instabilities. By taking the analytical $T \to 0$ limit (assuming a finite classical $C_V$), thermal kinetics are suppressed, isolating the geometric response driven strictly by cohesive potentials. In this limit, both $\bm{g}$ and $\bm{a}$ exhibit simultaneous $\mathcal{O}(T^{-2})$ rank reductions. Within their shared null space, the primary algebraic divergences elegantly cancel out~\cite{SM}, subjecting the slow mode to the subleading negative curvature of the interaction potential
\begin{equation}\label{eq:negative_curvature}
\lambda_{\mathrm{s}} = \min_{\bm{v}} \frac{\bm{v}^\top \bm{g} \bm{v}}{\bm{v}^\top \bm{a} \bm{v}} \approx -\beta \Xi < 0.
\end{equation}
This emergence of a negative relaxation rate rigorously signals macroscopic structural instability, such as spinodal decomposition (e.g., spinodal-triggered gelation~\cite{luGelationParticlesShortrange2008}). It showcases the framework's ability to systematically detect non-equilibrium structural arrests via algebraic singularity tracking.

\textit{Concluding remarks}.---In summary, our thermodynamic geometry framework for relaxation can be intuitively understood through a simple mechanical analogy: the local relaxation rate of a 1D overdamped particle is dictated by the ratio of its restoring potential curvature to its viscous drag. Elevating this scalar ratio to a multidimensional state space, we recast intrinsic thermodynamic evolution as a tensorial competition between the thermodynamic potential curvature ($\bm{g}$) and the thermodynamic friction ($\bm{a}$). We demonstrated that a system's intrinsic relaxation rates and principal dynamic axes emerge naturally from this geometric quotient as coordinate-independent invariants. Applying this framework to the van der Waals fluid, we reveal the scaling laws of macroscopic asymptotic dynamics and the geometric origin of critical slowing down. 

Furthermore, this framework establishes a versatile foundation for broader nonequilibrium explorations. First, it corroborates the temporal decomposition of thermodynamic metrics~\cite{liDecompositionMetricTensor2025} and advances standard sloppiness analysis~\cite{brownStatisticalMechanicalApproaches2003, transtrumGeometryNonlinearLeast2011, machtaParameterSpaceCompression2013, transtrumPerspectiveSloppinessEmergent2015} by revealing the dynamic geometric origin of macroscopic dimensionality reduction. Additionally, exploring the relaxation geometry of driven phase transitions~\cite{2026Chen} presents an exciting avenue for investigating dynamic scaling around phase transition point. Second, embedding this intrinsic eigenspectrum into dynamic renormalization group flows~\cite{hohenbergTheoryDynamicCritical1977, machtaParameterSpaceCompression2013, maityInformationGeometryRenormalization2015} provides a systematic pathway to incorporate multiscale fluctuations and extract anomalous dynamic exponents. Third, mapping this eigenspectrum promises to unveil the origins of anomalous behaviors, such as the Mpemba effect~\citep{lu_2017,lin2026macroscopic} and its quantum counterparts~\citep{Chatterjee_2023_PRL,ares2025quantum}, where temporal hierarchies are counterintuitively inverted. Finally, the universality of this approach naturally extends to stochastic and quantum thermodynamics~\citep{Ito2018,Quantum-Geo-2020,LiGengGeodesicPath2022,2024Zhao}. Reformulating this geometric ratio via the quantum Fisher information metric and the Lindbladian dissipation matrix could decode decoherence and entanglement dynamics, ultimately unifying complex relaxation phenomena within a single geometric paradigm.

	\makeatletter
	\let\oldaddcontentsline\addcontentsline
	\renewcommand{\addcontentsline}[3]{} 
	\makeatother
	
\textit{Acknowledgments}.---We thank Y. Q. Lin, R. H. Chen, W. C. Xu, and K. T. Huang for insightful discussions. This work is supported by the Fundamental and Interdisciplinary Disciplines Breakthrough Plan of the Ministry of Education of China (Grant No. JYB2025XDXM305). Y.-H.M. acknowledges support from the National Natural Science Foundation of China under Grant No. 12305037.
	
\bibliography{Refs}
	
	\makeatletter
	\let\addcontentsline\oldaddcontentsline
	\makeatother

	
\clearpage 
\onecolumngrid 

\thispagestyle{empty}
\begin{center}
	\textbf{\large Supplemental Material for ``Thermodynamic Geometry of Relaxation''}\\[0.5cm]
	
	Hao Wang$^{1,\star}$(wang\_hao1@tju.edu.cn), Li Zhao$^{1}$, Shuai Deng$^{1}$ and Yu-Han Ma$^{2,3,\dagger}$(yhma@bnu.edu.cn)\\[0.3cm]
	
{\small
		$^1$\textit{State Key Laboratory of Engines, Tianjin University, Tianjin 300072, China}\\
		$^2$\textit{School of Physics and Astronomy, Beijing Normal University, Beijing, 100875, China}\\
		$^3$\textit{Key Laboratory of Multiscale Spin Physics (Beijing Normal University), Ministry of Education, Beijing 100875, China}
	}

\end{center}

\vspace{0.5cm}

This Supplemental Material provides detailed mathematical derivations, exact analytical limits, and comprehensive parameter settings to support the findings presented in the main text. The outline of the supporting materials are listed as follows

\setcounter{equation}{0}
\setcounter{figure}{0}
\setcounter{table}{0}
\setcounter{page}{1}
\setcounter{section}{0}    
\setcounter{subsection}{0} 
\setcounter{secnumdepth}{3}
\makeatletter
\renewcommand{\theequation}{S\arabic{equation}}
\renewcommand{\thefigure}{S\arabic{figure}}
\renewcommand{\thetable}{S\arabic{table}}
\renewcommand{\thesection}{\Roman{section}}
\renewcommand{\thesubsection}{\Alph{subsection}}
\renewcommand{\theHsection}{S\Roman{section}}
\renewcommand{\theHsubsection}{S\Alph{subsection}}
\renewcommand{\theHequation}{S\arabic{equation}}
\renewcommand{\theHfigure}{S\arabic{figure}}
\renewcommand{\theHtable}{S\arabic{table}}
\makeatother

\addtocontents{toc}{\protect\setcounter{tocdepth}{3}}
\vspace{0.4cm}
\noindent\rule{\textwidth}{0.4pt}
\vspace{-0.2cm}
\tableofcontents
\vspace{0.2cm}
\noindent\rule{\textwidth}{0.4pt}
\vspace{0.8cm}

\section{Macroscopic Thermodynamic Geometric Framework}

\begin{figure}[h]
	\centering
	\includegraphics[width=0.4\linewidth]{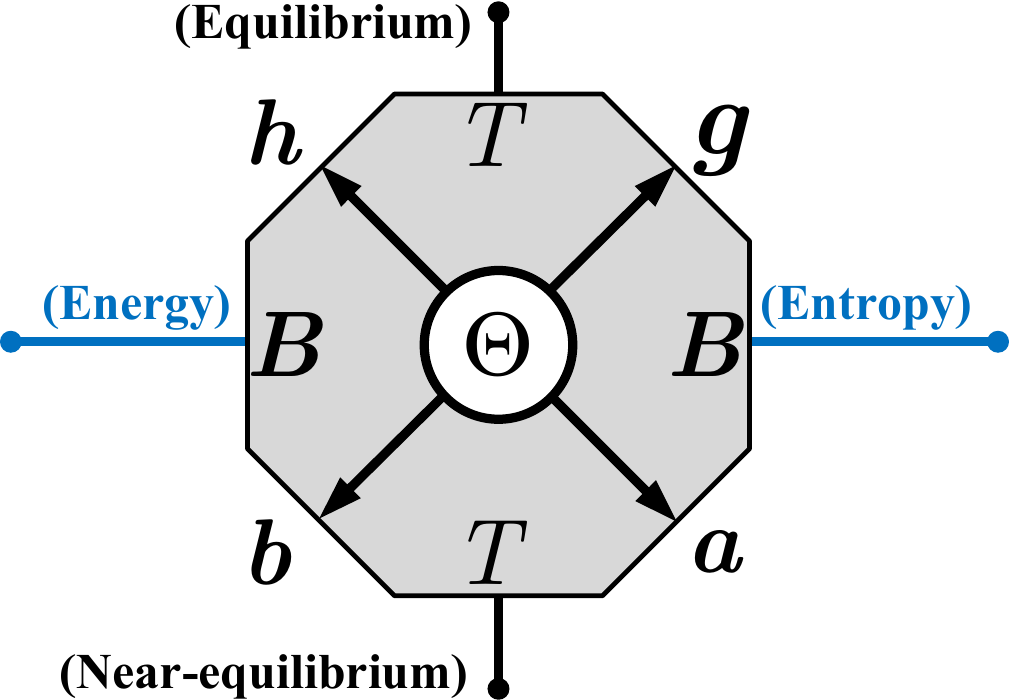}
	\caption{Covariant architecture of macroscopic thermodynamic geometry. Temperature $T$ acts as a conformal factor connecting the entropy ($\bm{g}, \bm{a}$) and energy ($\bm{h}, \bm{b}$) representations. The relaxation tensor $\bm{B}$ acts as a dynamical distortion operator mapping static equilibrium geometry onto near-equilibrium dissipative geometry. Commutativity strictly ensures representation-independent intrinsic dynamical modes.}
	\label{fig:relations_of_metrics}
\end{figure}

While the main text relies on the entropy representation to highlight entropy-driven relaxation, we extend this geometric framework here to the energy representation, constructing a covariant unified architecture comprising four metrics (Fig.~\ref{fig:relations_of_metrics}). This establishes theoretical self-consistency and identifies the relaxation time tensor $\bm{B}$ as a universal geometric operator bridging thermodynamic potentials and transport processes.

The metrics (Fig.~\ref{fig:relations_of_metrics}) are classified by physical representation (energy/entropy) and regime (static/dynamic). Defining extensive variables $\bm{X} \in \mathbb{R}^n$ and state deviations $\bm{x} = \bm{X} - \bm{X}_{\mathrm{eq}}$, the static geometry comprises the Ruppeiner metric $\bm{g} = -\nabla^2 S$ and Weinhold metric $\bm{h} = \nabla^2 U$~\cite{ruppeinerThermodynamicsRiemannianGeometric1979b, weinholdMetricGeometryEquilibrium1975}. Concurrently, the dynamic dissipative geometry defines the generalized friction $\bm{a}$ via the entropy production rate $\dot{\sigma} = \dot{\bm{x}}^\top \bm{a} \dot{\bm{x}}$~\cite{Sprikl1995}, and the Rayleigh dissipation metric $\bm{b}$ via the energy dissipation function $\Phi = \dot{\bm{x}}^\top \bm{b} \dot{\bm{x}}$~\cite{Goldstein2002}. 

Governed by $\Phi = T \dot{\sigma}$ and the conformal transformation $\bm{h}=T\bm{g}$~\cite{salamonRelationEntropyEnergy1984}, these metrics form a strictly commutative diagram:
\begin{align}
	\text{Representation Transformation:} & \quad \bm{h} = T \bm{g}, \quad \bm{b} = T \bm{a} \label{eq:Horizontal_metrics}\\ 
	\text{Dynamical Mapping:} & \quad \bm{b} = \bm{h} \bm{B}, \quad \bm{a} = \bm{g} \bm{B}. \label{eq:Vertical_metrics}
\end{align}
This algebraic closure ensures framework covariance. The relaxation tensor $\bm{B}$ identically maps static restorative stiffness to kinetic friction ($\bm{a} = \bm{g}\bm{B}$ or $\bm{b} = \bm{h}\bm{B}$), independent of the chosen representation.

\subsection{Thermodynamic Normal Modes and Simultaneous Diagonalization}

The relaxation rate matrix is defined as:
\begin{equation}\label{eq:Gamma_B}
\bm{\Gamma} = \bm{B}^{-1} = \bm{a}^{-1}\bm{g}.
\end{equation}
Constrained by Onsager reciprocity and Maxwell relations, $\bm{\Gamma}$ exhibits generalized self-adjointness: $\bm{\Gamma}^\top \bm{g} = \bm{g}\bm{\Gamma}$. 

This symmetry guarantees real eigenvalues and enables simultaneous diagonalization of the static and dynamic metrics. Decomposing $\bm{\Gamma} = \bm{Q} \bm{\Lambda} \bm{Q}^{-1}$, with eigenvalue matrix $\bm{\Lambda} = \operatorname{diag}(\lambda_\mathrm{f}, \lambda_\mathrm{s})$ and principal axes $\bm{Q} = (\bm{v}_\mathrm{f}, \bm{v}_\mathrm{s})$, the linear phenomenological equation $\dot{\bm{x}}=-\bm{\Gamma}\bm{x}$ diagonalizes into the normal mode representation:
\begin{equation} \label{eq:eq_decoupled}
\dot{\hat{\bm{x}}}=-\bm{\Lambda} \hat{\bm{x}}, \quad \text{where} \quad \hat{\bm{x}}=\bm{Q}^{-1}\bm{x}.
\end{equation}
Consequently, multivariable relaxation decouples into pure exponential decay modes $\hat{x}_i(t)\propto e^{-\lambda_i t}$. In this intrinsic basis, tensorial mappings reduce to scalar scaling relations for each principal mode $i$: 
\begin{equation}
\tilde{b}_{i} = \tau_i \tilde{h}_{i} = T \tau_i \tilde{g}_{i} = T \tilde{a}_{i},
\end{equation}
where the intrinsic relaxation time $\tau_i = 1/\lambda_i$ serves as a local geometric scaling factor coupling static curvature to the dynamic dissipative background.

\section{Detailed Derivations from the Variational Principle to the Rayleigh Quotient}

To extract intrinsic evolutionary modes from the state space geometry, we define the thermodynamic Rayleigh quotient for any non-zero $\bm{u} \in \mathbb{R}^n$:
\begin{equation}
\mathcal{R}(\bm{u}) \equiv \frac{\bm{u}^\top \bm{g} \bm{u}}{\bm{u}^\top \bm{a} \bm{u}}. \label{eq:Rayleigh_Def}
\end{equation}
Imposing the stationarity condition $\delta \mathcal{R} = 0$ directly yields the generalized eigenvalue problem $\bm{g}\bm{v} = \lambda \bm{a}\bm{v}$. Thus, the stationary directions of $\mathcal{R}(\bm{u})$ identically correspond to the system's principal axes $\bm{v}$, with extremal values given by the eigenvalues $\lambda$.

To anchor this geometric construction to non-equilibrium kinetics, we employ the dynamical Galerkin weak form~\cite{finlaysonMethodWeightedResiduals1972}. Defining the instantaneous residual $\bm{r}(t) = \bm{g}\bm{x} + \bm{a}\dot{\bm{x}} = \bm{0}$ and introducing the single-mode ansatz $\bm{x}(t) = \bm{u} e^{-\lambda t}$, we construct the action functional over the trajectory $t \in [0, \infty)$:
\begin{equation} \label{eq:SM_functional}
\Pi(\bm{u}, \lambda) = \int_0^\infty \bm{x}^\top (\bm{g}\bm{x} + \bm{a}\dot{\bm{x}}) dt.
\end{equation}
Time integration yields the explicit functional:
\begin{equation} \label{eq:SM_functional_explicit}
\Pi = \frac{1}{2\lambda} \left( \bm{u}^\top \bm{g} \bm{u} - \lambda \bm{u}^\top \bm{a} \bm{u} \right).
\end{equation}
The Ritz stationarity principle, $\partial \Pi / \partial \bm{u} = \bm{0}$, requires $\bm{g}\bm{u} - \lambda \bm{a}\bm{u} = \bm{0}$. Hence, macroscopic relaxation trajectories inherently lock to the generalized eigenvectors $\bm{v}$. Left-multiplying by $\bm{v}^\top$ identifies the physical relaxation rate as the stationary thermodynamic Rayleigh quotient: $\lambda = \mathcal{R}(\bm{v})$.

\section{Dynamical Model And Thermodynamic Metrics}

\subsection{Mechanical Dissipation Model in the Overdamped Limit}
The non-equilibrium evolution model of the van der Waals argon gas consists of thermal conduction (Newton's cooling, heat transfer coefficient $\alpha$) and mechanical damping. Consider a piston-dashpot system where $m$ is the piston mass, $A$ is its cross-sectional area, and $x$ is its position. Newton's second law dictates:
\begin{equation}\label{eq:piston_motion}
m \ddot{x} = (P - P_{\text{ext}})A - \gamma \dot{x},
\end{equation}
where $\gamma$ is the linear mechanical damping coefficient. In the macroscopic regime on thermodynamic relaxation timescales, inertial effects are negligible compared to the pressure gradient and viscous drag. Employing the overdamped approximation ($m\ddot{x} \approx 0$), the macroscopic volume rate $\dot{V} = A\dot{x}$ exacts to:
\begin{equation}
\dot{V} = \frac{A^2}{\gamma} (P - P_{\text{ext}}).
\end{equation}
Defining the phenomenological mechanical mobility as $\beta \equiv A^2/\gamma$, we formally obtain the linear kinetic equations $\dot{Q} = \alpha(T_{\mathrm{ext}} - T)$ and $\dot{V} = \beta(P - P_{\text{ext}})$.

\subsection{Construction of the State Space Metrics}
The system evolves strictly within the internal energy-volume $(U, V)$ state space. For a van der Waals fluid, the equation of state is explicitly given by
\begin{equation}\label{eq:vdW_EOS}
P = \frac{RT}{V-b} - \frac{a}{V^2}.
\end{equation}
Correspondingly, the fundamental equation in the entropy representation is formulated as:
\begin{equation}\label{eq:entropy}
S(U, V) = R \ln(V-b) + C_V \ln\left(U + \frac{a}{V}\right) + \text{const}.
\end{equation}

The thermodynamic metric in the Ruppeiner geometry is defined by the negative Hessian of the entropy density with respect to the state variables, $g_{ij} = -\partial^2 S / \partial x^i \partial x^j$ with $x^i \in \{U, V\}$. Evaluating the first-order derivatives of Eq.~(\ref{eq:entropy}) yields the thermal and mechanical equations of state:
\begin{align}
	\frac{\partial S}{\partial U} &= \frac{C_V}{U + a/V} = \frac{1}{T}, \\
	\frac{\partial S}{\partial V} &= \frac{R}{V-b} - \frac{C_V a}{V^2(U + a/V)} = \frac{P}{T}.
\end{align}
Applying the second-order partial derivatives and substituting the thermal relation $U + a/V = C_V T$ directly isolates the metric components:
\begin{align}
	g_{UU} &= -\frac{\partial^2 S}{\partial U^2} = \frac{C_V}{(U + a/V)^2} = \frac{1}{C_V T^2}, \\
	g_{UV} &= -\frac{\partial^2 S}{\partial U \partial V} = -\frac{a}{C_V T^2 V^2}, \\
	g_{VV} &= -\frac{\partial^2 S}{\partial V^2} = \frac{R}{(V-b)^2} + \frac{a^2}{C_V T^2 V^4} - \frac{2a}{T V^3}.
\end{align}

Incorporating the van der Waals parameters ($a, b$) along with the ideal isochoric heat capacity ($C_V$), the exact analytical Ruppeiner static metric is:
\begin{equation}\label{eq:Van_Ruppeiner}
\bm{g} = \frac{1}{T^2} 
\begin{pmatrix} 
\frac{1}{C_V} & \chi \\ 
\chi & \frac{a^2}{C_V V^4} - T \Xi 
\end{pmatrix}.
\end{equation}

Here, the cross-coupling term $\chi$ and the macroscopic isothermal volumetric stiffness $\Xi$ are defined as:
\begin{equation}\label{eq:SM_chi}
\chi \equiv \left(\frac{\partial T}{\partial V}\right)_U = -\frac{a}{C_V V^2},
\end{equation}
\begin{equation}\label{eq:SM_auxiliary_stiffness}
\Xi \equiv \left(\frac{\partial P}{\partial V}\right)_T = -\frac{RT}{(V-b)^2} + \frac{2a}{V^3}.
\end{equation}

To explicitly derive the Onsager transport matrix $\bm{L}$ satisfying $\bm{J} = \bm{L} \Delta \bm{F}$, we choose fluxes $\bm{J} = (\dot{U}, \dot{V})^\top$ and conjugate thermodynamic forces in the entropy representation $\Delta \bm{F} = \nabla S \approx (\Delta(1/T), \Delta(P/T))^\top$. Through a first-order Taylor expansion near equilibrium, we map the measurable variations $(\delta T, \delta P)$ to the forces:
\begin{align}
	\Delta F_1 &\approx -\frac{1}{T^2}\delta T \implies \delta T \approx -T^2 \Delta F_1, \\
	\Delta F_2 &\approx \frac{1}{T}\delta P - \frac{P}{T^2}\delta T \implies \delta P \approx T \Delta F_2 - PT \Delta F_1.
\end{align}
Substituting this mapping into the phenomenological equations ($\dot{Q} = -\alpha \delta T$, $\dot{V} = \beta \delta P$) and combining it with the first law of thermodynamics ($\dot{U} = \dot{Q} - P\dot{V}$), we deduce the fluxes explicitly in terms of $\Delta \bm{F}$:
\begin{align}
	J_2 &= \dot{V} = -\beta P T \Delta F_1 + \beta T \Delta F_2, \\
	J_1 &= \dot{U} = (\alpha T^2 + \beta P^2 T)\Delta F_1 - \beta P T \Delta F_2.
\end{align}
Extracting these coefficients formulates the Onsager matrix $\bm{L}$. The symmetry $L_{12} = L_{21} = -\beta P T$ rigorously validates Onsager reciprocity mechanically. The generalized dissipation metric $\bm{a}$ is identically its inverse:
\begin{equation}\label{eq:SM_a}
\bm{L} = \begin{pmatrix} \alpha T^2 + \beta P^2 T & -\beta P T \\-\beta P T & \beta T \end{pmatrix} \implies \bm{a} = \bm{L}^{-1} = \begin{pmatrix}\frac{1}{\alpha T^2} & \frac{P}{\alpha T^2} \\ \frac{P}{\alpha T^2} & \frac{1}{\beta T} + \frac{P^2}{\alpha T^2}\end{pmatrix}.
\end{equation}

Table~\ref{tab:parameters} summarizes the phenomenological settings used for numerical verification. Suppressing the mechanical mobility $\beta$ artificially induces a strongly coupled, frictionally dominated regime optimal for isolating the underlying metric competition.

\begin{table}[h]
	\caption{Summary of phenomenological constants, thermodynamic fluid parameters (Argon), and kinetic coefficients utilized for numerical verification.}
	\label{tab:parameters}
	\renewcommand{\arraystretch}{1.5}
	\begin{ruledtabular}
		\begin{tabular}{ll|ll}
			\textbf{Parameter} & \textbf{Value} & \textbf{Parameter} & \textbf{Value} \\
			\colrule
			vdW attraction $a$ & $0.1363\,\mathrm{Pa \cdot m^6 / mol^2}$ & Heat transfer coeff. $\alpha$ & $7.0 \,\mathrm{W / K}$ \\
			vdW covolume $b$ & $3.219 \times 10^{-5}\,\mathrm{m^3 / mol}$ & Mechanical mobility $\beta$ & $6.0 \times 10^{-8}\,\mathrm{m^3 / (Pa \cdot s)}$ \\
			Isochoric capacity $C_V$ & $1.5 R \approx 12.47\,\mathrm{J / (mol\cdot K)}$ & Equilibrium Temp. $T_{\mathrm{eq}}$ & $300\,\mathrm{K}$ \\
			Universal gas const. $R$ & $8.314\,\mathrm{J / (mol\cdot K)}$ & Equilibrium Press. $P_{\mathrm{eq}}$ & $2 \times 10^5\,\mathrm{Pa}$ \\
		\end{tabular}
	\end{ruledtabular}
\end{table}

\section{Asymptotic Behaviors of Characteristic Relaxation Rates}

Within the phenomenological framework, the state space is spanned by the increments $\bm{x} = (\Delta U, \Delta V)^\top$. To circumvent analytical singularities arising from the divergence of the variational denominator under extreme conditions, we utilize the metrics $\bm{a}$ and $\bm{g}$. The fast and slow relaxation rates, $\lambda_{\mathrm{f}}$ and $\lambda_{\mathrm{s}}$, correspond directly to the eigenvalues of $\bm{\Gamma}$ as defined in Eq.~\eqref{eq:Gamma_B}.

In the following derivations, calculating the determinant of $\bm{\Gamma}$ algebraically verifies the rigidity constraints generating the scaling laws. Concurrently, by extracting the degenerate forms of $\bm{g}$ and $\bm{a}$ at various asymptotic limits, we geometrically elucidate the manifold locking phenomena summarized in Table~\ref{tab:relaxation_rates}.

\subsection{Auxiliary Thermodynamic Relations and Incremental Forms}

Based on the mechanical equation of state in Eq.~(\ref{eq:vdW_EOS}) and the explicit definitions of $\chi$ and $\Xi$ given in Eqs.~(\ref{eq:SM_chi}) and (\ref{eq:SM_auxiliary_stiffness}), we extract the internal pressure $\pi$ and its thermodynamic identity connecting to the isochoric parameter:
\begin{equation}\label{eq:SM_internal_pressure}
\pi \equiv \left(\frac{\partial U}{\partial V}\right)_T = \frac{a}{V^2}, \quad C_V \chi = -\pi.
\end{equation}
Consequently, the first-order thermodynamic increments governing the internal energy and reversible heat exchange are formulated as:
\begin{equation}\label{eq:SM_increments}
\Delta U = C_V \Delta T + \pi \Delta V, \quad \Delta Q = \Delta U + P \Delta V.
\end{equation}
By substituting the internal pressure relation $\chi = -\pi/C_V$ from Eq.~\eqref{eq:SM_internal_pressure}, the full relaxation rate matrix defined in Eq.~\eqref{eq:Gamma_B} can be calculated as:

\begin{equation} \label{eq:rate_matrix}
\bm{\Gamma} = 
\begin{pmatrix}
\frac{\alpha}{C_V} + \frac{\beta P(P+\pi)}{C_V T} & -\frac{\alpha \pi}{C_V} + \beta P \Xi - \frac{\beta P \pi (P+\pi)}{C_V T} \\
-\frac{\beta(P+\pi)}{C_V T} & -\beta \Xi + \frac{\beta \pi (P+\pi)}{C_V T}
\end{pmatrix}.
\end{equation}
The term $(P+\pi)$ represents the total effective pressure driving mechanical relaxation. To isolate the mechanisms governing these dynamics, we analyze the structural degeneration of $\bm{\Gamma}$ under extreme macroscopic conditions. Specifically, we decompose this matrix to reveal how limits in pressure ($P$, see \ref{sec:Pressure}) and temperature ($\epsilon$, $T$,  see \ref{sec:epsilon} and \ref{sec:Tto0}) dictate asymptotic geometric behaviors and timescale separations.

\subsection{Pressure Asymptotics}\label{sec:Pressure}

With the temperature fixed at $T=300\,\mathrm{K}$, we treat pressure $P$ as the sole control parameter. To establish a global scaling constraint, we calculate the metric determinants, noting the exact cancellation of the off-diagonal terms:
\begin{align}
	\det(\bm{a}^{-1}) &= (\alpha T^2 + \beta P^2 T)(\beta T) - (-\beta P T)^2 = \alpha \beta T^3, \\
	\det(\bm{g}) &= \frac{1}{C_V T^2}\left(-\frac{\Xi}{T} + \frac{C_V \chi^2}{T^2}\right) - \left(\frac{\chi}{T^2}\right)^2 = -\frac{\Xi}{C_V T^3}.
\end{align}
Following the properties of $\bm{\Gamma} = \bm{a}^{-1}\bm{g}$ established in Eq.~\eqref{eq:Gamma_B}, the product of the characteristic rates $\lambda_{\mathrm{f}} \lambda_{\mathrm{s}}$ identically equals $\det(\bm{\Gamma})$. Simplifying the temperature factor $T^3$ yields a global invariant:
\begin{equation}\label{eq:global_invariant_SM}
\lambda_{\mathrm{f}} \lambda_{\mathrm{s}} = \det(\bm{\Gamma}) = \det(\bm{a}^{-1})\det(\bm{g}) = -\frac{\alpha \beta \Xi}{C_V}.
\end{equation}
In both the low-pressure ($V \approx RT/P$) and high-pressure hard-core ($V-b \approx RT/P$) limits, the isothermal volumetric stiffness degenerates to $\Xi \approx -P^2/(RT)$. Substituting this into Eq.~\eqref{eq:global_invariant_SM} reveals a volumetric rigidity constraint valid across the entire parameter space:
\begin{equation}\label{eq:scaling_law_pressure}
\lambda_{\mathrm{f}} \lambda_{\mathrm{s}} \approx \frac{\alpha \beta P^2}{C_V R T} \propto P^2.
\end{equation}

This quadratic scaling law bounds the coupled relaxation rates by macroscopic volumetric rigidity. Driven by this global invariant, thermodynamic geometry undergoes fundamentally distinct degeneration and manifold locking at pressure extremes. We unfold these derivations below for the dilute ($P \to 0$, see \ref{subsubsection:deltaV_0}) and dense ($P \to \infty$, see \ref{sec:high_p_pole}) limits.

\subsubsection{Low-Pressure Limit ($P \to 0$)}\label{subsubsection:deltaV_0}

Under the asymptotic conditions of the dilute gas limit ($V \to \infty$), the equation of state reduces to the ideal gas law ($V \approx RT/P$) from Eq.~\eqref{eq:vdW_EOS}. Consequently, the parameter degeneration leads to the decay of internal pressure and attractive terms ($\pi \to 0$, $\chi \to 0$) as derived from Eqs.~\eqref{eq:SM_chi} and \eqref{eq:SM_internal_pressure}, while the macroscopic volumetric stiffness decays algebraically ($\Xi \approx -P^2/(RT) \to 0$) from Eq.~\eqref{eq:SM_auxiliary_stiffness}.

Under these limits, the static metric $\bm{g}$ in Eq.~\eqref{eq:Van_Ruppeiner} degenerates to a rank-1 matrix, geometrically indicating a complete loss of restorative force along the volumetric dimension $(0, 1)^\top$:
\begin{equation}
\lim_{P \to 0} \bm{g} \approx 
\begin{pmatrix} 
\frac{1}{C_V T^2} & 0 \\ 
0 & 0 
\end{pmatrix} = 
\frac{1}{C_V T^2} \begin{pmatrix} 1 \\ 0 \end{pmatrix} (1, 0).
\end{equation}

Driven by the decay of cross-terms, decoupling emerges as the matrix in Eq.~\eqref{eq:rate_matrix} explicitly diagonalizes:
\begin{equation}
\bm{\Gamma} \approx 
\begin{pmatrix} 
\frac{\alpha}{C_V} & 0 \\ 
0 & \frac{\beta P^2}{R T} 
\end{pmatrix}.
\end{equation}
This directly yields the decoupled characteristic rates:
\begin{align}
	\lambda_{\mathrm{f}} &= \frac{\alpha}{C_V}, \label{eq:lambda_low_p_fast} \\
	\lambda_{\mathrm{s}} &= \frac{\beta P^2}{R T} \propto P^2 \to 0. \label{eq:lambda_low_p_slow}
\end{align}

For the fast mode $\lambda_{\mathrm{f}}$, the eigenvalue equation $(\bm{\Gamma} - \lambda_{\mathrm{f}} \bm{I})\bm{v}_{\mathrm{f}} = \bm{0}$ requires:
\begin{equation}
\begin{pmatrix} 
0 & 0 \\ 
0 & \frac{\beta P^2}{R T} - \frac{\alpha}{C_V} 
\end{pmatrix} 
\begin{pmatrix} \Delta U \\ \Delta V \end{pmatrix} = \begin{pmatrix} 0 \\ 0 \end{pmatrix}.
\end{equation}
Since $\frac{\beta P^2}{R T} \to 0$ and $\frac{\alpha}{C_V} \neq 0$, this algebraically forces $\Delta V = 0$, explicitly locking the fast evolution onto the isochoric manifold:
\begin{equation}\label{eq:manifold_low_p_fast}
\Delta V = 0 \quad (\text{Isochoric manifold}).
\end{equation}

Conversely, the slow mode aligns with $(0,1)^\top$, requiring $\Delta U = 0$ (isoenergetic manifold). With the internal pressure vanished ($\pi \to 0$), Eq.~\eqref{eq:SM_increments} simplifies to $\Delta U \approx C_V \Delta T$. Therefore, isoenergetic evolution here is strictly equivalent to the isothermal manifold:
\begin{equation}\label{eq:manifold_low_p_slow}
\Delta T = 0 \quad (\text{Isothermal manifold}).
\end{equation}

\subsubsection{High-Pressure Limit ($P \to \infty$)} \label{sec:high_p_pole}

In the dense hard-core compression physical limit ($V \to b$), the equation of state is dominated by repulsion $P \approx RT/(V-b)$, yielding $(\partial P/\partial T)_V \approx P/T$ from Eq.~\eqref{eq:vdW_EOS}. This causes a parameter degeneration where the volumetric stiffness diverges ($\Xi \approx -P^2/(RT) \to -\infty$) from Eq.~\eqref{eq:SM_auxiliary_stiffness}, and the macroscopic pressure dominates the internal pressure ($P \gg \pi$) from Eq.~\eqref{eq:SM_internal_pressure}.

As $P \to \infty$, the resistance metric $\bm{a}$ in Eq.~\eqref{eq:SM_a} partitions by divergence order into thermal and mechanical components:
\begin{equation}
\bm{a} = \frac{1}{\alpha T^2} 
\begin{pmatrix} 1 & P \\ P & P^2 \end{pmatrix} + 
\begin{pmatrix} 0 & 0 \\ 0 & \frac{1}{\beta T} \end{pmatrix}.
\end{equation}
Extracting the principal $\mathcal{O}(P^2)$ divergent term yields the degenerate outer product:
\begin{equation}
\bm{a}^{\mathrm{div}} = \frac{1}{\alpha T^2} \begin{pmatrix} 1 \\ P \end{pmatrix} (1, P).
\end{equation}
To bypass this divergent resistance, the fast mode trajectory must align with the null space of $\bm{a}^{\mathrm{div}}$:
\begin{equation}
(1, P) \begin{pmatrix} \Delta U \\ \Delta V \end{pmatrix} = 0 \implies \Delta U + P \Delta V = 0.
\end{equation}
Substituting this relation into Eq.~\eqref{eq:SM_increments} identifies the strict locking onto the adiabatic manifold:
\begin{equation}\label{eq:manifold_high_p_fast}
\Delta Q = 0 \quad (\text{Adiabatic manifold}).
\end{equation}

Applying the high-pressure limit approximations ($P \gg \pi \implies P+\pi \approx P$ and $\Xi \approx -P^2/(RT)$) to Eq.~\eqref{eq:rate_matrix} and retaining only the highest-order terms in $P$ yields the asymptotic matrix:
\begin{align}
	\bm{\Gamma} &\approx 
	\begin{pmatrix}
		\frac{\beta P^2}{C_V T} & \beta P \left(-\frac{P^2}{RT}\right) \\
		-\frac{\beta P}{C_V T} & -\beta \left(-\frac{P^2}{RT}\right)
	\end{pmatrix} \nonumber \\
	&= \frac{\beta P}{T} 
	\begin{pmatrix} 
		\frac{P}{C_V} & -\frac{P^2}{R} \\ 
		-\frac{1}{C_V} & \frac{P}{R} 
	\end{pmatrix} \equiv \bm{\Gamma}_0. \label{eq:Gamma_0}
\end{align}
Here, the off-diagonal coupling term $\Gamma_{12}$ diverges as $\mathcal{O}(P^3)$, dominating the asymptotic eigen-structure.

As $\det(\bm{\Gamma}_0) = 0$, its sole non-zero eigenvalue (the fast rate) equals its trace. Under the asymptotic limits $(\partial P/\partial T)_V \approx P/T$ and $\Xi \approx -P^2/(RT)$, the generalized Mayer's relation $C_P - C_V = -T (\partial P/\partial T)_V^2 / \Xi$ reduces to the ideal gas form:
\begin{equation}
C_P - C_V \approx -T \frac{(P/T)^2}{-P^2 / RT} = R.
\end{equation}
Consequently, the fast relaxation rate diverges quadratically:
\begin{equation}\label{eq:lambda_high_p_fast}
\lambda_{\mathrm{f}} \approx \operatorname{Tr}(\bm{\Gamma}_0) = \frac{\beta P^2}{C_V T} + \frac{\beta P^2}{R T} = \frac{\beta P^2 (R + C_V)}{C_V R T} \approx \frac{\beta P^2 C_P}{C_V R T} \propto P^2.
\end{equation}

The corresponding eigenvector $\bm{v}_{\mathrm{f}}$ spans the column space of $\bm{\Gamma}_0$, giving $(\Delta U, \Delta V)^\top \propto (P, -1)^\top$. Applying Eq.~\eqref{eq:SM_increments} verifies the geometric adiabatic constraint: $\Delta Q = \Delta U + P \Delta V = P + P(-1) = 0$.

The slow mode resides in the null space of $\bm{\Gamma}_0$, satisfying $\bm{\Gamma}_0 \bm{v}_{\mathrm{s}} = \bm{0} \implies R \Delta U = C_V P \Delta V$. With $P \gg \pi$, the increment relation in Eq.~\eqref{eq:SM_increments} reduces to $\Delta U \approx C_V \Delta T$, yielding:
\begin{equation}
R(C_V \Delta T) \approx C_V P \Delta V \implies R \Delta T \approx P \Delta V.
\end{equation}
Substituting this into the perturbed equation of state $P \Delta V + (V-b)\Delta P \approx R \Delta T$ forces $(V-b)\Delta P \approx 0$, constraining the slow mode strictly to the isobaric manifold:
\begin{equation}\label{eq:manifold_high_p_slow}
\Delta P = 0 \quad (\text{Isobaric manifold}).
\end{equation}
To extract the constant baseline rate of the slow mode without full perturbative expansion, we divide the global invariant from Eq.~\eqref{eq:scaling_law_pressure} by the fast mode rate $\lambda_{\mathrm{f}}$:
\begin{equation}\label{eq:lambda_high_p_slow}
\lambda_{\mathrm{s}} = \frac{\det(\bm{\Gamma})}{\lambda_{\mathrm{f}}} \approx \frac{\frac{\alpha \beta P^2}{C_V R T}}{\frac{\beta P^2 C_P}{C_V R T}} = \frac{\alpha}{C_P}.
\end{equation}

\subsubsection{Numerical Verification of Pressure Asymptotics}

\begin{figure}[htbp]
	\centering
	\includegraphics[width=0.5\linewidth]{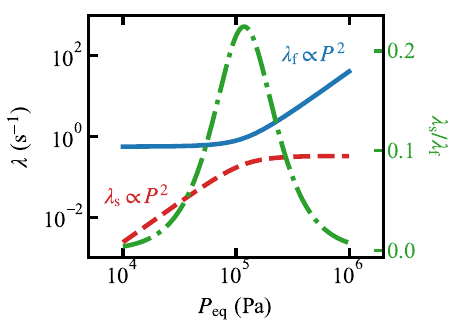} 
	\caption{Global dynamics of relaxation rates spanning the pressure limits. The numerical bounds ($10^4\,\mathrm{Pa}$ and $10^6\,\mathrm{Pa}$) correctly capture the analytically derived scalings ($\propto P^2$) and saturation plateaus. The right axis (green dash-dotted curve) evaluates the timescale ratio $\lambda_{\mathrm{s}}/\lambda_{\mathrm{f}}$, rigorously peaking at the kinetic crossover threshold $P_\times \approx 1.87 \times 10^5 \, \mathrm{Pa}$.}
	\label{fig:pressure_scaling}
\end{figure}

To numerically validate the derived scaling laws, we evaluate the exact relaxation rates across a wide pressure range (Fig.~\ref{fig:pressure_scaling}). From Eqs.~\eqref{eq:Van_Ruppeiner} and \eqref{eq:SM_a}, equating the diagonal components of $\bm{a}^{-1}$ defines a kinetic crossover threshold $P_\times = \sqrt{\alpha T/\beta} \approx 1.87 \times 10^5 \, \mathrm{Pa}$ (Table~\ref{tab:parameters}). This scale intrinsically bounds our numerical evaluation domain: the mathematical limits $P \to 0$ and $P \to \infty$ are effectively captured by the simulated dilute ($10^4\,\mathrm{Pa} \ll P_\times$) and dense ($10^6\,\mathrm{Pa} \gg P_\times$) regimes, respectively.

The numerical landscape corroborates the analytical limits. In the dilute regime, the numerical rates confirm the fast-mode plateau and quadratic slow-mode decay ($\lambda_{\mathrm{s}} \propto P^2$) derived in Eqs.~\eqref{eq:lambda_low_p_fast} and \eqref{eq:lambda_low_p_slow}. Conversely, at the dense extreme, the geometric mechanism inverts, validating the quadratic fast-mode divergence ($\lambda_{\mathrm{f}} \propto P^2$) and slow-mode saturation plateau established in Eqs.~\eqref{eq:lambda_high_p_fast} and \eqref{eq:lambda_high_p_slow}.

Furthermore, the timescale separation ratio $\lambda_{\mathrm{s}}/\lambda_{\mathrm{f}}$ in Fig.~\ref{fig:pressure_scaling} naturally peaks at $P_\times$. As discussed in the main text, this crossover marks the minimum timescale separation where the intrinsic anisotropies of $\bm{g}$ and $\bm{a}$ optimally counterbalance, temporarily blurring the distinct slow manifold and maximizing thermomechanical dissipation coupling.

\subsection{Reduced Temperature Asymptotics}\label{sec:epsilon}
Along the critical isochore ($V_{\mathrm{c}} = 3b$), we define the reduced temperature as $\epsilon = (T - T_{\mathrm{c}})/T_{\mathrm{c}} > 0$.
The isothermal volumetric stiffness $\Xi$ defined in Eq.~\eqref{eq:SM_auxiliary_stiffness}, when evaluated at $V_{\mathrm{c}} = 3b$, is:
\begin{equation}
\Xi(V_{\mathrm{c}}, T)  = -\frac{RT}{4b^2} + \frac{2a}{27b^3}.
\end{equation}
Substituting $T = T_{\mathrm{c}}(1 + \epsilon)$ and the critical temperature $T_{\mathrm{c}} = \frac{8a}{27bR}$ yields a linear dependence on $\epsilon$:
\begin{equation}
\Xi =  -\frac{2a}{27b^3} \epsilon.
\end{equation}
Using the global rate invariant from Eq.~\eqref{eq:global_invariant_SM}, this linear scaling simplifies the constraint to:
\begin{equation}\label{eq:scaling_law_temp}
\lambda_{\mathrm{f}} \lambda_{\mathrm{s}} = \left( \frac{2a \alpha \beta}{27 b^3 C_V} \right) \epsilon \propto \epsilon^1.
\end{equation} 

This linear scaling law indicates that along the critical isochore, evolutionary speed is governed by the thermal deviation from criticality. Under this constraint, the state space geometry exhibits contrasting asymptotic behaviors at the critical ($\epsilon \to 0$) and high-temperature ($\epsilon \to \infty$) limits, which we examine below.

\subsubsection{Critical Limit ($\epsilon \to 0$)}

Approaching the critical limit constrained to the critical isochore ($V_{\mathrm{c}} = 3b$) as $\epsilon \to 0$, we operate under the mean-field assumption that the isochoric heat capacity $C_V$ remains finite. In this regime, the parameters degenerate such that the volumetric stiffness vanishes ($\Xi \to 0$) as per Eq.~\eqref{eq:SM_auxiliary_stiffness}, and the internal pressure remains constant ($\pi \to \pi_{\mathrm{c}} = 3P_{\mathrm{c}}$) according to Eq.~\eqref{eq:SM_internal_pressure}.

The evolution rate at the critical point is directly given by the trace of $\bm{\Gamma}$ from Eq.~\eqref{eq:trace_gamma}:
\begin{equation}\label{eq:trace_gamma}
\operatorname{Tr}(\bm{\Gamma}) = \Gamma_{11} + \Gamma_{22} = \frac{\alpha}{C_V} - \beta \Xi + \frac{\beta}{C_V T}(P + \pi)^2.
\end{equation}

As $\Xi \to 0$, the trace converges to a strictly positive constant. Since the trace dictates the sum of the eigenvalues, the fast mode rate remains finite:
\begin{equation}\label{eq:lambda_crit_fast}
\lambda_{\mathrm{f}} \approx \operatorname{Tr}(\bm{\Gamma}) \to \text{constant}.
\end{equation}
Simultaneously, the slow mode rate isolates the linear scaling of the global invariant in Eq.~\eqref{eq:scaling_law_temp}:
\begin{equation}\label{eq:lambda_crit_slow}
\lambda_{\mathrm{s}} \approx \frac{\det(\bm{\Gamma})}{\operatorname{Tr}(\bm{\Gamma})} \propto \epsilon^1 \to 0.
\end{equation}

Using $C_V \chi = -\pi_{\mathrm{c}}$ from Eq.~\eqref{eq:SM_internal_pressure}, the critical static metric degenerates into a rank-1 outer product:
\begin{equation}
\bm{g}(T_{\mathrm{c}}) = 
\frac{1}{C_V T_{\mathrm{c}}^2} \begin{pmatrix} 1 \\ -\pi_{\mathrm{c}} \end{pmatrix} (1, -\pi_{\mathrm{c}}).
\end{equation}
Since $\bm{a}$ remains full rank, the vanishing slow mode ($\lambda_{\mathrm{s}} \to 0$) forces its eigenvector $\bm{v}_{\mathrm{s}} = (\Delta U, \Delta V)^\top$ into the null space of $\bm{g}(T_{\mathrm{c}})$:
\begin{equation}
(1, -\pi_{\mathrm{c}}) \begin{pmatrix} \Delta U \\ \Delta V \end{pmatrix} = 0 \implies \Delta U - \pi_{\mathrm{c}} \Delta V = 0.
\end{equation}
Substituting the internal energy increment $\Delta U = C_V \Delta T + \pi_{\mathrm{c}} \Delta V$ from Eq.~\eqref{eq:SM_increments} yields:
\begin{equation}\label{eq:delta_T_critical}
(C_V \Delta T + \pi_{\mathrm{c}} \Delta V) - \pi_{\mathrm{c}} \Delta V = C_V \Delta T = 0.
\end{equation}
Thus, critical slowing down is strictly locked onto the isothermal manifold:
\begin{equation}\label{eq:manifold_crit_slow}
\Delta T = 0 \quad (\text{Isothermal manifold}).
\end{equation}

Conversely, the fast mode eigenvector $\bm{v}_{\mathrm{f}}$ must be proportional to $\bm{a}^{-1} \bm{u}$, where $\bm{u} = (1, -\pi_{\mathrm{c}})^\top$. Substituting $\pi_{\mathrm{c}} = 3P_{\mathrm{c}}$ yields:
\begin{equation} 
\bm{v}_{\mathrm{f}} \propto \bm{a}^{-1} \begin{pmatrix} 1 \\ -3P_{\mathrm{c}} \end{pmatrix} = \begin{pmatrix} \alpha T_{\mathrm{c}}^2 + \beta P_{\mathrm{c}}^2 T_{\mathrm{c}} & -\beta P_{\mathrm{c}} T_{\mathrm{c}} \\ -\beta P_{\mathrm{c}} T_{\mathrm{c}} & \beta T_{\mathrm{c}} \end{pmatrix} \begin{pmatrix} 1 \\ -3P_{\mathrm{c}} \end{pmatrix}. 
\end{equation}
Extracting the volumetric component $\Delta V_{\mathrm{f}}$ (the second row) gives:
\begin{equation}\label{eq:locking_crit_slow} 
\Delta V_{\mathrm{f}} \propto (-\beta P_{\mathrm{c}} T_{\mathrm{c}}) \cdot 1 + (\beta T_{\mathrm{c}}) \cdot (-3P_{\mathrm{c}}) = -4\beta P_{\mathrm{c}} T_{\mathrm{c}}.
\end{equation}
Since $\beta, P_{\mathrm{c}}, T_{\mathrm{c}} > 0$, we find $\Delta V_{\mathrm{f}} \neq 0$. Therefore, unlike the strictly isochoric behavior in the low-pressure limit (see Sec.~\ref{subsubsection:deltaV_0}), the volumetric component of the critical fast mode converges to a finite, non-zero constant, explicitly demonstrating a thermo-mechanical mixed evolution (see Sec.~\ref{sec:manifold_locking}).

\subsubsection{High-Temperature Limit ($\epsilon \to \infty$)}

Operating within the ideal gas physical limit ($\epsilon \to \infty$), the macroscopic pressure diverges ($P \propto \epsilon \to \infty$), reducing Mayer's relation to $C_P - C_V = R$. Due to parameter degeneration, the volumetric stiffness diverges ($-\Xi \approx P^2/(RT) \propto \epsilon^1$) according to Eq.~\eqref{eq:SM_auxiliary_stiffness}, and the attractive internal pressure vanishes ($\pi \to 0$, $\chi \to 0$) as per Eq.~\eqref{eq:SM_internal_pressure}.

With $\pi \to 0$ and $P^2 \approx RT(-\Xi)$, the quadratic term in the trace formula of Eq.~\eqref{eq:trace_gamma} simplifies to $(P + \pi)^2 \approx P^2$. Neglecting lower-order constants and substituting the ideal gas Mayer's relation, the trace diverges linearly:
\begin{align}
	\operatorname{Tr}(\bm{\Gamma}) &\approx -\beta \Xi + \frac{\beta P^2}{C_V T} \approx -\beta \Xi \left( 1 + \frac{R}{C_V} \right) \nonumber \\
	&\approx -\beta \Xi \left( \frac{C_P}{C_V} \right) \propto \epsilon^1.
\end{align}

Because the trace diverges ($\mathcal{O}(\epsilon)$), it entirely dictates the fast mode. Utilizing the linear global invariant constraint in Eq.~\eqref{eq:scaling_law_temp}, the characteristic rates decouple into standalone limits:
\begin{equation}\label{eq:lambda_high_temp_fast}
\lambda_{\mathrm{f}} \approx \operatorname{Tr}(\bm{\Gamma}) \propto \epsilon^1 \to \infty,
\end{equation}
\begin{equation}\label{eq:lambda_high_temp_slow}
\lambda_{\mathrm{s}} \approx \frac{\det(\bm{\Gamma})}{\operatorname{Tr}(\bm{\Gamma})} \propto \frac{\epsilon^1}{\epsilon^1} \to \text{constant}.
\end{equation}

By directly applying the parameter degenerations ($\pi \to 0$ and $\Xi \approx -P^2/(RT)$) to the full evolution matrix $\bm{\Gamma}$ in Eq.~\eqref{eq:rate_matrix} and extracting the dominant divergent terms, we yield the asymptotic evolution matrix:
\begin{align}
	\bm{\Gamma} &\approx 
	\begin{pmatrix} 
		\frac{\beta P^2}{C_V T} & \beta P \left(-\frac{P^2}{RT}\right) \\ 
		-\frac{\beta P}{C_V T} & -\beta \left(-\frac{P^2}{RT}\right) 
	\end{pmatrix} \nonumber \\
	&= \frac{\beta P}{T} 
	\begin{pmatrix} 
		\frac{P}{C_V} & -\frac{P^2}{R} \\ 
		-\frac{1}{C_V} & \frac{P}{R} 
	\end{pmatrix} \equiv \bm{\Gamma}_0.
\end{align}

This high-temperature asymptotic matrix $\bm{\Gamma}_0$ is algebraically identical to the divergent matrix governing the extreme high-pressure limit shown in Eq.~\eqref{eq:Gamma_0}. This isomorphism confirms that both macroscopic extremes trigger an identical matrix degeneration driven by divergent dissipative resistance. 

Consequently, the high-temperature eigenvectors perfectly match the high-pressure constraints. The diverging fast mode ($\lambda_{\mathrm{f}} \propto P^2/T \propto \epsilon^2/\epsilon = \epsilon^1$) evades the resistance by locking strictly onto the adiabatic manifold established in Eq.~\eqref{eq:manifold_high_p_fast}:
\begin{equation}\label{eq:manifold_high_temp_fast}
\Delta Q = 0 \quad (\text{Adiabatic manifold}).
\end{equation}
Simultaneously, the slow mode eigenvector falls into the null space of $\bm{\Gamma}_0$, locking onto the isobaric manifold from Eq.~\eqref{eq:manifold_high_p_slow}:
\begin{equation}\label{eq:manifold_high_temp_slow}
\Delta P = 0 \quad (\text{Isobaric manifold}).
\end{equation}

\subsection{Absolute Zero Limit ($T \to 0$)}\label{sec:Tto0}

Taking the $T \to 0$ physical limit under the classical assumption of constant $C_V$ to isolate intermolecular attractive effects, the repulsive thermal pressure vanishes ($RT/(V-b) \to 0$) according to Eq.~\eqref{eq:vdW_EOS}. This parameter degeneration causes the macroscopic pressure to reduce to the attractive internal pressure ($P \to P_0 \equiv -\pi =C_V \chi= -a/V^2$) based on Eqs.~\eqref{eq:vdW_EOS} and \eqref{eq:SM_internal_pressure}, leaving the macroscopic stiffness purely attractive ($\Xi \approx 2a/V^3 > 0$) as seen in Eq.~\eqref{eq:SM_auxiliary_stiffness}.

By applying the parameter degenerations $P \to P_0$ and $\pi \to -P_0$ to the full kinetic matrix $\bm{\Gamma}$ in Eq.~\eqref{eq:rate_matrix}, the effective mechanical driving pressure perfectly vanishes ($P + \pi \to 0$). This exact cancellation eliminates the thermal divergent terms and the lower-left matrix element, directly reducing the evolution matrix to a strict upper triangular form:
\begin{equation} \label{eq:gamma_upper_triangular}
\lim_{T \to 0}\bm{\Gamma} = 
\begin{pmatrix} 
\frac{\alpha}{C_V} & P_0 \left( \frac{\alpha}{C_V} + \beta \Xi \right) \\ 
0 & -\beta \Xi 
\end{pmatrix}.
\end{equation}
The characteristic fast and slow evolution rates correspond identically to the main diagonal elements:
\begin{equation}\label{eq:lambda_T0_fast}
\lambda_{\mathrm{f}} = \frac{\alpha}{C_V},
\end{equation}

\begin{equation}\label{eq:lambda_T0_slow}
\lambda_{\mathrm{s}} = -\beta \Xi.
\end{equation}

Solving the eigenvectors of this asymptotic upper triangular matrix explicitly resolves the physical constraints on the evolution manifolds. For the fast mode $\lambda_{\mathrm{f}}$, the characteristic equation $(\bm{\Gamma} - \lambda_{\mathrm{f}} \bm{I})\bm{v}_{\mathrm{f}} = \bm{0}$ leverages the zero lower-left element to strictly force the volumetric component to zero, locking the evolution onto the isochoric manifold:
\begin{equation}\label{eq:manifold_T0_fast}
\Delta V = 0 \quad (\text{Isochoric manifold}).
\end{equation}
Conversely, the slow mode $\lambda_{\mathrm{s}}$ yields the top-row constraint $\Delta U + P_0 \Delta V = 0$. Substituting $P_0 = -\pi$ and the internal energy increment $\Delta U = C_V \Delta T + \pi \Delta V$ from Eq.~\eqref{eq:SM_increments}, the volume increments perfectly cancel ($C_V \Delta T = 0$), strictly constraining the slow mode trajectory to the isothermal manifold:
\begin{equation}\label{eq:manifold_T0_slow}
\Delta T = 0 \quad (\text{Isothermal manifold}).
\end{equation}

The algebraic upper-triangularization fundamentally originates from the concurrent degeneration of the foundational metrics. Utilizing the identity $C_V \chi = P_0=-a/V^2$, the static metric $\bm{g}$ in Eq.~\eqref{eq:Van_Ruppeiner} and the dissipation metric $\bm{a}$ in Eq.~\eqref{eq:SM_a} are decomposed by their $T$-divergence order:
\begin{align}
	\bm{g} &= \frac{1}{C_V T^2} \begin{pmatrix} 1 \\ P_0 \end{pmatrix} (1, P_0) + \begin{pmatrix} 0 & 0 \\ 0 & -\frac{\Xi}{T} \end{pmatrix}, \label{eq:g_split} \\
	\bm{a} &\approx \frac{1}{\alpha T^2} \begin{pmatrix} 1 \\ P_0 \end{pmatrix} (1, P_0) +\begin{pmatrix} 0 & 0 \\ 0 & \frac{1}{\beta T} \end{pmatrix}. \label{eq:a_split}
\end{align}
As $T \to 0$, the $\mathcal{O}(T^{-2})$ dominant divergent terms in both matrices reduce to rank-1 forms that precisely align in the state space, pointing along the identical direction $(1, P_0)^\top$. 
This geometric alignment guarantees the boundedness of the fast mode rate. While the infinite restoring force and immense resistance both diverge as $\mathcal{O}(T^{-2})$ along this direction, they exactly cancel each other out within the thermodynamic Rayleigh quotient defined in Eq.~\eqref{eq:Rayleigh_Def}:
\begin{equation}
\mathcal{R}(\bm{v}_{\mathrm{f}}) \approx \frac{\frac{1}{C_V T^2} [\bm{v}_{\mathrm{f}}^\top (1, P_0)^\top]^2}{\frac{1}{\alpha T^2} [\bm{v}_{\mathrm{f}}^\top (1, P_0)^\top]^2} \xrightarrow{T \to 0} \frac{\alpha}{C_V} \equiv \lambda_{\mathrm{f}}.
\end{equation}

To evade the $\mathcal{O}(T^{-2})$ divergent resistance barrier, the slow mode $\bm{v}_{\mathrm{s}}$ inevitably retreats into the orthogonal null space (the isothermal manifold). Having escaped the dominant divergent terms of both metrics, the system evolution is exposed solely to the secondary static topology:
\begin{equation}
\bm{v}_{\mathrm{s}}^\top \bm{g} \bm{v}_{\mathrm{s}} = 1^2 \cdot \left(-\frac{\Xi}{T}\right) = -\frac{\Xi}{T}.
\end{equation}
Because the intermolecular attraction strictly dominates the macroscopic stiffness at this limit ($\Xi \approx 2a/V^3 > 0$), the effective restorative force inverts to a negative value. This geometric inversion establishes a saddle-point topological constraint:
\begin{equation}\label{eq:saddle_point_curvature}
-\frac{\Xi}{T} < 0 \quad (\text{Saddle-point negative curvature}).
\end{equation}

\begin{table}[htbp]
	\caption{Summary of global evolution and manifold locking of characteristic relaxation rates.}
	\label{tab:relaxation_rates}
	\renewcommand{\arraystretch}{1.5} 
	\begin{ruledtabular}
		\begin{tabular}{llll}
			\textbf{Physical Limit} & \textbf{Mode} & \textbf{Rate ($\lambda$)} & \textbf{Constraint \& Features} \\
			\colrule
			\multicolumn{4}{c}{\textbf{Control Parameter: Pressure $P$}} \\
			\colrule
			Low $P \to 0$ & Fast $\lambda_{\mathrm{f}}$ & $\frac{\alpha}{C_V}$~\eqref{eq:lambda_low_p_fast} & 
			\parbox[c]{7.5cm}{\vspace{0.15cm} \raggedright $\Delta V = 0$~\eqref{eq:manifold_low_p_fast} (isochoric); static metric degeneration aligns with the non-zero force direction. \vspace{0.15cm}} \\
			
			& Slow $\lambda_{\mathrm{s}}$ & $\propto P^2 \to 0$~\eqref{eq:lambda_low_p_slow} & 
			\parbox[c]{7.5cm}{\vspace{0.15cm} \raggedright $\Delta T = 0$~\eqref{eq:manifold_low_p_slow} (isothermal); loss of volume stiffness falls into the static metric null space. \vspace{0.15cm}} \\
			\colrule
			High $P \to \infty$ & Fast $\lambda_{\mathrm{f}}$ & $\propto P^2 \to \infty$~\eqref{eq:lambda_high_p_fast} & 
			\parbox[c]{7.5cm}{\vspace{0.15cm} \raggedright $\Delta Q = 0$~\eqref{eq:manifold_high_p_fast} (adiabatic); dissipation metric degeneration escapes along the zero-resistance direction. \vspace{0.15cm}} \\
			
			& Slow $\lambda_{\mathrm{s}}$ & $\frac{\alpha}{C_P}$~\eqref{eq:lambda_high_p_slow} & 
			\parbox[c]{7.5cm}{\vspace{0.15cm} \raggedright $\Delta P = 0$~\eqref{eq:manifold_high_p_slow} (isobaric); evades resistance divergence, finding sanctuary. \vspace{0.15cm}} \\
			\colrule
			\multicolumn{4}{c}{\textbf{Control Parameter: Temperature $T$ ($\epsilon$)}} \\
			\colrule
			Critical $\epsilon \to 0$ & Fast $\lambda_{\mathrm{f}}$ & (const.)~\eqref{eq:lambda_crit_fast} & 
			\parbox[c]{7.5cm}{\vspace{0.15cm} \raggedright Trace converges smoothly, maintaining a constant baseline \eqref{eq:locking_crit_slow}. \vspace{0.15cm}} \\
			
			& Slow $\lambda_{\mathrm{s}}$ & $\propto \epsilon^1 \to 0$~\eqref{eq:lambda_crit_slow} & 
			\parbox[c]{7.5cm}{\vspace{0.15cm} \raggedright $\Delta T = 0$~\eqref{eq:manifold_crit_slow} (isothermal); static metric degeneration leads to a loss of restorative force. \vspace{0.15cm}} \\
			\colrule
			High $\epsilon \to \infty$ & Fast $\lambda_{\mathrm{f}}$ & $\propto \epsilon^1 \to \infty$~\eqref{eq:lambda_high_temp_fast} & 
			\parbox[c]{7.5cm}{\vspace{0.15cm} \raggedright $\Delta Q = 0$~\eqref{eq:manifold_high_temp_fast} (adiabatic); due to underlying isomorphism, dissipation diverges globally. \vspace{0.15cm}} \\
			
			& Slow $\lambda_{\mathrm{s}}$ & (const.)~\eqref{eq:lambda_high_temp_slow} & 
			\parbox[c]{7.5cm}{\vspace{0.15cm} \raggedright $\Delta P = 0$~\eqref{eq:manifold_high_temp_slow} (isobaric); rigorously isomorphic to the high-pressure constant limit. \vspace{0.15cm}} \\
			\colrule
			Absolute Zero $T \to 0$ & Fast $\lambda_{\mathrm{f}}$ & $\frac{\alpha}{C_V}$~\eqref{eq:lambda_T0_fast} & 
			\parbox[c]{7.5cm}{\vspace{0.15cm} \raggedright $\Delta V = 0$~\eqref{eq:manifold_T0_fast} (isochoric); divergent terms cancel out; matrix forms an upper triangle. \vspace{0.15cm}} \\
			
			& Slow $\lambda_{\mathrm{s}}$ & $-\beta \Xi < 0$~\eqref{eq:lambda_T0_slow} & 
			\parbox[c]{7.5cm}{\vspace{0.15cm} \raggedright $\Delta T = 0$~\eqref{eq:manifold_T0_slow} (isothermal); saddle-point negative curvature~\eqref{eq:saddle_point_curvature} causes instability. \vspace{0.15cm}} \\
		\end{tabular}
	\end{ruledtabular}
\end{table}

\vspace{0.5cm}

\subsection{The Ideal Gas Limit: Absence of Critical Scaling}

To analytically verify that the geometric rank reduction strictly originates from cohesive interactions, we evaluate the non-interacting ideal gas limit ($a = 0, b = 0$). The fundamental equation strictly decouples, leading to vanishing cross-coupling ($\chi = 0$) and zero internal pressure ($\pi = 0$). Consequently, the static Ruppeiner metric formulated in Eq.~\eqref{eq:Van_Ruppeiner} degenerates into a strictly diagonal tensor
\begin{equation}
\bm{g}_{\mathrm{id}} = 
\begin{pmatrix} 
\frac{1}{C_V T^2} & 0 \\ 
0 & \frac{R}{V^2} 
\end{pmatrix}.
\end{equation}
Integrating the dynamic dissipation metric $\bm{a}$ from Eq.~\eqref{eq:SM_a}, the global rate identity evaluates to:
\begin{equation}
\lambda_{\mathrm{f}} \lambda_{\mathrm{s}} = \det(\bm{a}^{-1})\det(\bm{g}_{\mathrm{id}}) = \frac{\alpha \beta R T}{C_V V^2}.
\end{equation}
For any finite macroscopic state ($T > 0, V < \infty$), this product is strictly positive ($\lambda_{\mathrm{f}} \lambda_{\mathrm{s}} > 0$). Both metrics inherently maintain full rank, algebraically precluding geometric collapse. Therefore, anomalous critical scaling laws ($\lambda_{\mathrm{s}} \to 0$ at finite states) are strictly absent. The rate vanishes ($\lambda_{\mathrm{s}} \propto V^{-2} \to 0$) solely in the asymptotic vacuum limit ($V \to \infty$) due to trivial density depletion.

	\section{Geometric Landscapes of Extreme Limits}\label{sec:manifold_locking}
	To provide an intuitive physical picture, we visualize the thermodynamic Rayleigh quotient on the $(\Delta U, \Delta V)$ plane, demonstrating how extreme metric degenerations lock trajectories onto specific manifolds or drive spinodal divergence.

	\subsection{Manifold Locking of the Principal Axes}
	
	Taking the critical point ($\epsilon \to 0$) as an example, the geometric collapse of the static geometry ($\delta \mathcal{R} = 0$) compels the kinetic slow principal axis $\bm{v}_{\mathrm{s}}$ to align strictly with the static soft axis $\bm{v}_{\mathrm{g}}$ (the collapsing null space of $\bm{g}$). Crucially, this space identifies identically with the isothermal manifold ($\Delta T = 0$) defined in Eq.~\eqref{eq:manifold_crit_slow}.
	
	To quantify this geometric locking, we define the locking angle $\theta$:
	\begin{equation}
	\theta = \arccos \left( \frac{|\tilde{\bm{v}}_{\mathrm{s}}^\top \tilde{\bm{v}}_{\mathrm{g}}|}{\|\tilde{\bm{v}}_{\mathrm{s}}\| \|\tilde{\bm{v}}_{\mathrm{g}}\|} \right)
	\end{equation}
	This locking is visually quantified in Fig.~\ref{fig:combined_landscapes}(a). The scaling, $\theta \propto \epsilon$ as $\epsilon \to 0$ confirms that critical slowing down results from the kinetic bottleneck becoming strictly geometrically locked onto the collapsing isothermal direction, as expressed in Eq.~\eqref{eq:manifold_crit_slow}.
	
	\subsection{The Rayleigh Quotient Landscape and Thermo-Mechanical Coupling}
	
	Substituting the isothermal slow mode constraint ($\Delta T = 0$) defined in Eq.~\eqref{eq:manifold_crit_slow} into the exact internal energy increment relation $\Delta U = C_V \Delta T + \pi \Delta V$ from Eq.~\eqref{eq:SM_increments}, the thermal fluctuation term perfectly vanishes. This rigorously yields a coupled trajectory governed entirely by the critical internal pressure $\pi_{\mathrm{c}}$:
	\begin{equation}\label{eq:thermo_mech_coupling}
	\Delta U = \pi_{\mathrm{c}} \Delta V
	\end{equation}
	For argon fluid ($\pi_{\mathrm{c}} \approx 1.4 \times 10^7 \, \mathrm{Pa}$), this results in a trajectory with an exceptionally small slope on the $(\Delta U, \Delta V)$ phase plane:
	\begin{equation}
	k = \frac{\Delta V}{\Delta U} = \pi_{\mathrm{c}}^{-1} \approx 10^{-7} \, \mathrm{Pa}^{-1}
	\end{equation}
	
	Within the near-critical ($\epsilon=10^{-2}$) Rayleigh quotient landscape $\mathcal{R}(\bm{v})$ given by Eq.~\eqref{eq:Rayleigh_Def} (Fig.~\ref{fig:combined_landscapes}(b)), the phase trajectories exhibit distinct timescale separation: an initial relaxation governed by the fast mode $\bm{v}_{\mathrm{f}}$ (associated with large $\mathcal{R}$ and finite volumetric evolution described in Eq.~\eqref{eq:locking_crit_slow}), followed by an asymptotic decay along the slow mode $\bm{v}_{\mathrm{s}}$, which is strictly constrained to the isothermal manifold.
	
	\subsection{The Spinodal Instability Landscape near Absolute Zero}
	
	Extending the geometric framework to the absolute zero limit ($T \to 0$), intermolecular attraction strictly dominates, inverting the effective restorative force and yielding a negative slow mode rate ($\lambda_{\mathrm{s}} < 0$) as shown in Eq.~\eqref{eq:lambda_T0_slow}.
	
\begin{figure*}[htbp]
		\centering
		\includegraphics[width=\linewidth]{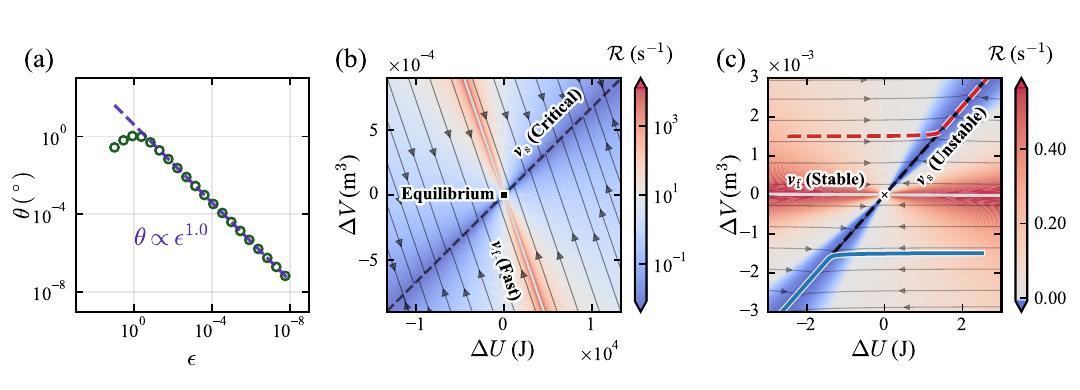} 
		\caption{Geometric landscapes of extreme limits. (a) Power-law decay of locking angle $\theta$ vs. reduced temperature $\epsilon$, quantifying the critical alignment ($\bm{v}_{\mathrm{s}} \to \bm{v}_{\mathrm{g}}$) driven by the structural collapse of static geometry. (b) Rayleigh quotient landscape near criticality ($\epsilon=10^{-2}$). Shown are the fast manifold ($\bm{v}_\mathrm{f}$, solid white, Eq.~\eqref{eq:locking_crit_slow}) and the slow manifold ($\bm{v}_\mathrm{s}$, dashed black) locked to the isothermal manifold (Eq.~\eqref{eq:manifold_crit_slow}). (c) Rayleigh quotient landscape near absolute zero ($T \to 0$). Shown are the stable fast manifold ($\bm{v}_\mathrm{f}$, dashed white) and the unstable spinodal valley ($\bm{v}_\mathrm{s}$, dashed black). Trajectories (solid blue, dashed red) illustrate macroscopic structural instability, diverging from the central equilibrium upon initial perturbation.}
		\label{fig:combined_landscapes}
	\end{figure*}

	To construct the landscape in Fig.~\ref{fig:combined_landscapes}(c), the macroscopic volume is fixed at $V = 0.0124\,\mathrm{m^3/mol}$, which corresponds to the equilibrium state at ambient conditions ($300\,\mathrm{K}$, $2 \times 10^5\,\mathrm{Pa}$). By instantly dropping the temperature inside this fixed volume, we remove the outward pressure of thermal motion, directly isolating the instability caused by intermolecular attraction.
	Within this landscape, the saddle-point negative curvature drives macroscopic structural instability: minute volume deviations are no longer restored, compelling dynamic trajectories to diverge from the central equilibrium strictly along the unstable slow mode ($\bm{v}_{\mathrm{s}}$).

\end{document}